\documentclass{kapedbk} 

\usepackage[dvips]{graphicx}

\chaptersection 

\upperandlowercase

 \let\footnote\savefootnote

\kluwerbib 

\def\mh{{M_{\bullet}}}
\def\mb{{M_{\rm bulge}}}
\def\rb{{\rho_{\bullet}}}
\def\mpc{{\rm Mpc}}

\def\lae{\mathrel{<\kern-1.0em\lower0.9ex\hbox{$\sim$}}}
\def\gae{\mathrel{>\kern-1.0em\lower0.9ex\hbox{$\sim$}}}

\newcommand{\msun}{{\rm M}_\odot}

\newcommand{\be}{\begin{equation} }
\newcommand{\ee}{\end{equation} }
\newcommand{\bea}{\begin{eqnarray} }
\newcommand{\eea}{\end{eqnarray} }

\begin{document}

\articletitle{Modeling the accretion history of supermassive black holes}
\rhead{Modeling the accretion history}

\author{Priyamvada Natarajan}
\affil{Department of Astronomy \& Department of Physics\\
Yale University, 260 Whitney Avenue, New Haven, CT 06511}
\email{priya@astro.yale.edu}

\begin{abstract}
There is overwhelming evidence for the presence of supermassive black
holes (SMBHs) in the centers of most nearby galaxies. The mass estimates 
for these remnant black holes from the stellar kinematics of
local galaxies and the quasar phenomenon at high redshifts
point to the presence of assembled SMBHs. The accretion history of 
SMBHs can be reconstructed using observations at high and low redshifts 
as model constraints. Observations of galaxies and quasars in the 
submillimeter, infrared, optical, and X-ray wavebands are used as 
constraints, along with data from the demography of local black holes.
Theoretical modeling of the growth of black hole mass with cosmic
time has been pursued thus far in two distinct directions: a
phenomenological approach that utilizes observations in various
wavebands, and a semi-analytic approach that starts with a theoretical
framework and a set of assumptions with a view to matching
observations. Both techniques have been pursued in the context of the
standard paradigm for structure formation in a Cold Dark Matter
dominated universe. In this chapter, we examine the key issues 
and uncertainties in the theoretical understanding of the growth 
of SMBHs. 
\end{abstract}


\section{Introduction}

The local demography of black holes (Ferrarese, this volume)
has established that most galaxies
harbor a supermassive black hole (SMBH; Kormendy \& Richstone 1995;
Magorrian et al.\ 1998; van der Marel 1999), most likely
assembled via a combination of accretion and mergers. These nuclear
SMBHs are ``dead quasars'', relics of quasar 
activity that might have occurred in many galaxies over their history
(Lynden-Bell 1969; So\l tan 1982; Rees 1990; Richstone et al.\ 1998). 
Early attempts to interlink the properties of these remnant
black holes with those of their host galaxy luminosities (Magorrian et
al.\ 1998) yielded a relation between the bulge luminosity and the black
hole mass of about 0.5 dex in the ratio of $\mh/\mb$.
A tighter correlation has since been measured between the
velocity dispersion of the bulge and the black hole mass (Ferrarese \&
Merritt 2000; Gebhardt et al.\ 2000), 
suggesting that the formation and evolution of SMBHs is inextricably 
linked to that of the stellar component of galactic bulges.

Quasar activity is powered by gas accretion onto SMBHs
(Lynden-Bell 1969), and hence the build-up of these SMBHs is likely
to have commenced at fairly high redshifts. Indeed, optically bright 
quasars have now been detected at redshifts greater than 6 
(e.g., Fan et al.\ 2001a, 2003). 
Quasars at these high redshifts provide an efficient tool to investigate 
the relation between black hole and early spheroid assembly. There are
indications that high redshift quasar hosts are often strong sources
of dust emission (Omont et al.\ 2001; Cox et al.\ 2002; 
Carilli et al.\ 2002; Walter et al.\ 2003), 
suggesting that quasars were common in massive galaxies at a time when 
the galaxies were undergoing copious star formation.

The growth of black hole mass in the universe can therefore be
traced using quasar activity. The phenomenological approach to
understanding the assembly of SMBHs involves using
observational data from both high and low redshifts as a starting point
to construct a viable and consistent picture that is consonant
with the larger framework of the growth and evolution of structure in
the universe. Another approach that has been pursued is
semi-analytic modeling, in which one starts from a set of ab initio 
assumptions and attempts to explain the observations. 

Both approaches have proved
to be fruitful and, in fact, share many common features. They
are both grounded in the framework of the standard paradigm that
involves the growth of structure via gravitational amplification of
small perturbations in a Cold Dark Matter (hereafter, CDM) universe---a
model that has independent validation, most recently from 
{\it Wilkinson Microwave Anisotropy Probe (WMAP)}
measurements of the anisotropies in the cosmic microwave background
(Spergel et al.\ 2003; Page et al.\ 2003). Structure formation in both
modeling schemes is tracked in cosmic time by keeping a census of the
number of collapsed dark matter halos of a given mass that form; these
provide the sites for harboring black holes. The computation of the
mass function of dark matter halos is done using either the
Press-Schechter (Press \& Schechter 1974) or the extended 
Press-Schechter theory (Lacey \& Cole 1993), or, in some cases, 
directly from cosmological N-body simulations (Di Matteo et al.\ 2003). 
In this chapter, we present the detailed modeling procedures
and key parameters and uncertainties for the
phenomenological approach, as discussed by Haehnelt, Natarajan, \& Rees
(1998). We also summarize the results from the semi-analytic modeling.

\section{The Phenomenological Approach: Basic Notation}

We first outline some of the basic definitions and assumptions 
that are useful in understanding the accretion paradigm. The 
Eddington luminosity of a black hole of mass $\mh$ is defined to be 
\bea 
L_{\rm Edd} = \frac{4\,\pi\,G\,c\,m_p\,\mh}{\sigma_T} \,, 
\eea 
where $m_p$ is the proton mass and $\sigma_T$ is
the Thomson scattering cross-section. The bolometric luminosity 
of the accreting black hole is given by 
\be 
L_{\rm bol} = \epsilon\,\dot M\,c^2 \,,
\ee
where $\dot M$ is the mass accretion rate and $\epsilon$ (typically 
assumed to be 10\%) is the radiative efficiency factor.
The Eddington rate is
defined to be the mass accretion rate for which a black hole with
radiative efficiency $\epsilon = 0.1$ has the Eddington luminosity,
\bea 
\dot M_{\rm Edd} = \frac{L_{\rm Edd}}{0.1~c^2} =
2.2~\Bigl(\frac{\mh}{10^8~\msun}\Bigr)~\msun~{\rm yr}^{-1} \,.
\eea 
The dimensionless rate $\dot m$ is simply the accretion rate
measured in units of the Eddington rate, 
$\dot m = \dot M/\dot M_{\rm Edd}$. 

This definition of the Eddington rate applies in the case of 
accretion onto a black hole from a thin accretion disk whose
viscosity $\nu = \alpha~c_s~H$ is defined in terms of the parameter
$\alpha$, the sound speed $c_s$, and the disk scale height $H$. 
The mass growth rate of a black hole accreting at 
$\dot M_{\rm Edd}$ is exponential with an e-folding 
timescale
\bea 
t_{\rm Salp} = 4.5 \times 10^7 {\rm yr} \,,
\eea 
the Salpeter time. For Eddington accretion, this is the only 
characteristic timescale in the problem.

\section{Observational Constraints from High and Low Redshifts}

In tying the various lines of observational evidence together, we
first consider the abundance and luminosity function
(LF) of quasars at various redshifts.
While bright quasar activity in the optical seems to peak around a
redshift of $z=2.5-3.0$ (Schmidt, Schneider, \& Gunn 1994; Warren,
Hewett, \& Osmer 1994; Shaver et al.\ 1996;
Boyle et al.\ 2000; Fan et al.\ 2001b), luminous quasars are now
detected at redshifts beyond 6 (Fan et al.\ 2001a, 2003). 
Since quasars are believed to be powered by accretion onto 
SMBHs at the centers of galaxies, a number of authors have linked the
changes in quasar activity to changes in the availability of fuel supply
at the centers of the host galaxies (Rees 1984; Cavaliere \& Szalay
1986; Wandel 1991; Small \& Blandford 1992; Haehnelt \& Rees
1993). Efstathiou \& Rees (1988) also recognized that the peak of quasar
activity coincides with the time when the first deep potential wells
assemble in plausible variants of hierarchical cosmogonies in CDM
models.  This enables the linking of the formation of the central
black holes with that of the dark matter halos in which the host
galaxies assemble.
 
The past few years have seen dramatic observational improvements in
the detection of galaxies and quasars at high redshifts, transforming 
our knowledge of galaxy and star formation in the high redshift universe 
(e.g., Madau, Pozzetti, \& Dickinson 1998). 
Now there are also far more extensive data on the demography of
SMBHs in nearby galaxies (Ferrarese, this volume) 
and on low level activity of Active Galactic Nuclei (AGN) in both the 
optical and X-ray wavebands (Narayan \& Yi 1995; Mushotzky, this volume). 
We next discuss some of the implications for the formation and evolution 
of AGN and attempt to combine the evidence from low and high redshift data.

\section{Local Demography of Black Holes}

The last few years have seen tremendous progress in establishing the
existence of SMBHs. There are now a number of
excellent cases (including that of our own Galaxy) where observations
strongly imply the presence of a relativistic potential well (Watson
\& Wallin 1994; Miyoshi et al.\ 1995; Genzel et al.\ 1997). 
Magorrian et al.\ (1998) first published about thirty
estimates for the masses of the putative black holes in the bulges of
nearby galaxies. They confirmed previous claims of a strong correlation 
between black hole and bulge mass (Kormendy \& Richstone 1995). 
Recent determinations of this relation (Merritt \& Ferrarese 2001) give
\bea
\frac{\mh}{\mb}\approx 10^{-2.9} \,.
\eea
However, considering the observational scatter, a 
mildly non-linear relation would probably also be consistent with the
data. We would further like to clarify that a linear relation between 
black hole and bulge mass does not necessarily imply a linear relation 
between black hole and halo mass, and, as we will argue later, a
non-linear relation might be more plausible.

Subsequently, a tighter correlation was found between the black hole 
mass and the velocity dispersion of the bulge 
(Ferrarese \& Merritt 2000; Gebhardt et al.\ 2000; Tremaine et al.\ 2002).
Combining that relation with the distribution function of velocity
dispersions for nearby galaxies from the SDSS (Sheth et al.\ 2003),
Yu \& Tremaine (2002) determined the local black hole mass density to
be
\bea
\rb(z=0)\simeq (2.5\pm0.4)~\times 
10^5\Bigl(\frac{h}{0.65}\Bigr)^2~\msun~\mpc^{-3} \,.  
\eea
Although this number is reasonably well determined, possible systematic 
errors in the black hole mass estimates imply that this number is still 
somewhat uncertain (van der Marel 1999; Valluri, Merritt, \& Emsellem 2004). 

Integrating the LF of optically bright quasars provides a secure lower bound
to the mass density acquired by black holes via accretion at high redshifts 
(So\l tan 1982; Chokshi \& Turner 1992). Yu \& Tremaine (2002) determined 
the integrated mass density of accreting black holes at $z = 0$ to be
\bea
\rb(z=0)=2.1\times 10^5~\msun~\mpc^{-3} \,,
\eea
for their preferred values of the bolometric correction, $f_B$, and 
accretion efficiency, $\epsilon$. Somewhat remarkably, given that we 
have assumed no contribution to black hole mass growth from obscured 
accretion, these numbers appear to be in agreement, within substantial 
uncertainties.  

\section{Relating the LFs of Star-forming Galaxies and
Quasars to CDM Halos}
                                                                                
\subsection{Lyman Break Galaxies}

Steidel and collaborators (Steidel \& Hamilton 1992;
Steidel et al.\ 1996; Giavalisco, Steidel, \& Macchetto 1996) 
exploited a technique (the so-called ``drop out'' or 
``Lyman break'' technique)
developed by Cowie et al.\ (1988) for picking out galaxies at high 
redshifts. The procedure is to make ultraviolet and optical observations 
in various wavebands and then see how bright the galaxies 
are in each of the wavebands. Steidel and collaborators were interested
in galaxies at redshifts of around three. At these redshifts, intrinsic 
absorption by both hot stars in the galaxies and surrounding neutral 
gas causes the galaxies to appear very faint---or 
``drop out''---in the observed ultraviolet waveband.
In this way, Steidel and his collaborators found many hundreds of 
star-forming galaxies at $2.5<z<4.5$. These galaxies, called Lyman 
Break Galaxies (LBGs), bear 
a close resemblance to local starburst galaxies. The abundance of
LBGs is roughly half that of $L>{L^*}$ present-day galaxies. 
There are, however, no secure direct dynamical mass estimates for 
the LBGs. The relation of the LBG masses to the rate of detected
star formation is becoming more clear with the discovery of strong
outflows and winds in many cases (Adelberger et al.\ 2003). Strong
clustering detected in the LBGs at $z\sim3$ leads
to an interpretation of these objects as the potential progenitors of
massive galaxies at the present epoch.

\subsection{High Redshift Quasars}

\begin{figure}[tbh]
\vskip -1.5cm
\vspace{-3.5cm}
\centerline{\includegraphics[width=\textwidth]{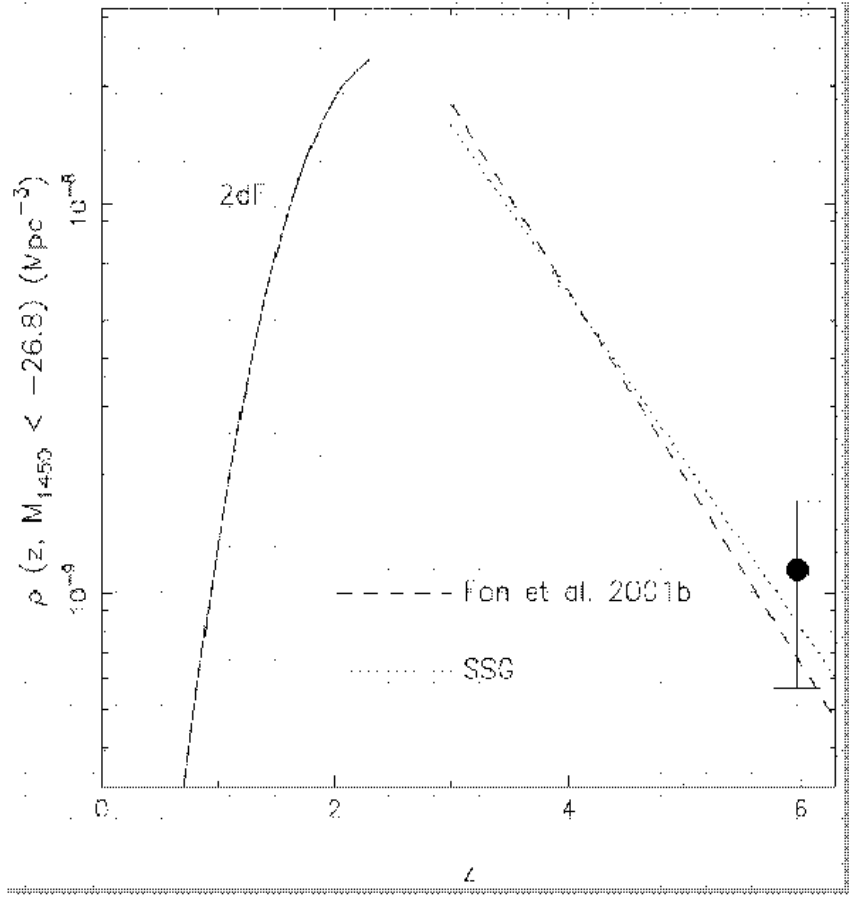}}
\caption{Evolution of the very luminous quasar comoving spatial
density at $M_B<-26.8$ in an Einstein-de Sitter universe with
$\Omega_M=1$ and $H_0=50$~km~s$^{-1}$~Mpc$^{-1}$. Large dot
represents the result from the SDSS (Fan et al.\ 2003).
Dashed and dotted lines are the best fit models from
Fan et al.\ (2001b) and Schmidt et al.\ (1994), respectively.
Solid line is the best fit model from the 2dF at $z<2.5$
(Boyle et al.\ 2000).
Figure from Fan et al.\ (2003; their Fig.~8).}
\label{priyafig1}
\end{figure}

Several groups have been engaged in the quest for high redshift
quasars. In the Palomar Transit Grism Survey,
Schmidt et al.\ (1994) detected 90 quasars in
the redshift range $z=2.75-4.75$ by their Ly$\alpha$ emission.
They found that the space density of
$M_B<-26$ quasars decreases by a factor of 2.7 per unit redshift
beyond $z=2.7$. Based on their analysis, they concluded that the 
peak of the comoving space density distribution of quasars with
$M_B<-26$ lies in the redshift range $z=1.7 - 2.7$. More recently, the
Anglo-Australian Telescope's Two Degree Field (2dF) redshift survey 
and the Sloan Digital Sky Survey
(SDSS) have detected quasars from $0<z<3$ (2dF) and beyond (SDSS).
Using color selection techniques, Fan et al.\ (2001b) derived the LF 
over the ranges $3.6<z<5.0$ and $-27.5<M_B<-25.5$ from a sample of
39 quasars from the SDSS; this LF is in good agreement with previous
estimates. Figure~\ref{priyafig1} shows how the space density of 
quasars detected at these bright magnitudes by the SDSS is consistent 
with extrapolations of the best fit LFs from the 2dF (Boyle et al.\ 2000).

Using their sample of six $z>6$ quasars, Fan et al.\ (2003) computed 
the space density of quasars at $z\sim 6$ to be
$\rho(M_B<-27.1)=(5\pm2)\times 10^{-10}$~Mpc$^{-3}$ for a
Lambda dominated universe with $\Omega_\Lambda=0.65$, $\Omega_M=0.35$,
and $H_0=65$~km~s$^{-1}$~Mpc$^{-1}$. The $z>6$ quasars have 
luminosities that imply black hole masses of a few times $10^9~\msun$, 
providing an important constraint for theoretical models.

%
%

The quasar LF at high redshifts also provides 
a sensitive test of cosmological parameters and models of quasar
evolution. The observed luminous quasars probably represent the rare
mass peaks in the density field at $z\sim 6$ and hence probe the
high mass tail of the dark halo mass function at these epochs. The
slope of the LF is determined by both the slope of the dark halo mass
function and the relation between black hole mass (proportional to
the quasar luminosity, if the quasar is radiating at the Eddington limit)
and dark halo mass. Since quasars are extremely rare at $z = 6$,
representing extremely high significance peaks in the underlying
density field, their LF is expected to be very steep. While current
determinations are not robust due to the small number of detected 
$z>6$ quasars, the slope at $z<2$ (Boyle et al.\ 2000) is found 
to be $\beta = -3.4$ (where the quasar LF is $\Psi (L) = L^{\beta}$), 
and the slope at $z \sim 4$ (Fan et al.\ 2001b) is found to be 
$\beta = -2.6$. An interesting fact to note here is that the mass 
function of CDM halos at $z=6$ is, in fact, steeper than the slope 
of the quasar LF at the relevant mass scale, which immediately 
suggests that for high redshift quasars, their luminosities cannot 
simply scale linearly with the mass of the halo $(M_{\rm halo})$ that 
hosts them; rather, high mass black holes radiate more efficiently 
than do lower mass ones.

All studies are in agreement with a steep decline in the abundance of 
quasars at low redshifts. At the high redshift end, the observed 
decline in the quasar number density could be due either to the real
paucity of quasars or to obscuration by dust, introducing a systematic
bias at high redshifts. Radio wavelengths are unaffected 
by dust: Shaver et al.\ (1996) report that in the Parkes radio
survey of a large sample of flat spectrum sources covering roughly 40\% 
of the sky, the space density of radio loud quasars does indeed decline
with redshift at $z>3$. These authors argue that the same conclusion
probably applies to all quasars.

\subsection{Linking Star-forming Galaxies and Quasars}

In what follows, we explore the link between star-forming galaxies and
quasars at high redshifts, assuming that both populations trace the 
mass function of DM halos in the universe. Using the Press-Schechter
formalism, we obtain estimates of the space density of DM halos for 
an assumed cosmological model. The space density of high redshift,
star-forming galaxies detected via the Lyman break technique
corresponds to that of halos with masses $\sim 10^{12.5}~\msun$ 
and virial velocities of order 300~km~s$^{-1}$ 
(see also Baugh, Frenk, \& Lacey\ 1998). 
Further evidence for masses of this order comes from the strong 
clustering of these galaxies (Steidel et al.\ 1998; Bagla 1998; 
Jing \& Suto 1998). 

A reasonable fit to 
the LF can be obtained by assuming a linear relation between the 
star formation rate and the halo mass, i.e., a constant
mass-to-light ratio. A weakly non-linear relation is also consistent
with the data and, indeed, would be required to match the shallow slope
of the LF at the high luminosity end, as reported by Bershady et al.\ (1997). 
Comparable fits are obtained for the other variants of the CDM model. 
The observed $H{\beta}$ widths of the LBGs of
$\sigma\sim 80$~km~s$^{-1}$ (Pettini et al.\ 1998) are 
assumed to be due to the massive amounts of star formation 
ongoing in the inner parts of the galaxies.
Nearby starbursts suggest that most of this emission originates
at very small radii, and hence these velocities do not reflect the 
virial velocities of $300$~km~s$^{-1}$ required from the inferred 
masses of CDM host halos.

The space density of optically selected quasars at $z=3$ with 
$M_B<-23$ is smaller than that of the detected star-forming galaxies by
a factor of a few hundred. As demonstrated by Haehnelt \& Rees (1993),
the evolution of optically selected quasars can be
linked to the hierarchical growth of DM halos on a similar timescale
only if the duration of the optically bright phase, $t_{\rm Q}$, is
considerably shorter than the Hubble time. For small $t_{\rm Q}$, this
is more and more in line with the predicted space densities of DM
halos and star-forming galaxies at high redshifts. However, hardly 
anything is known about the masses of the host galaxies of optically 
selected quasars, and this still leaves considerable freedom in
the exact choice of $t_{\rm Q}$. 

Following the approach of Haehnelt
et al.\ (1998), we estimate the formation rate of active black holes 
by taking the positive term of the time derivative of the
halo mass function and a simple parameterization for the black hole
formation efficiency. It is further assumed that active black holes
radiate with a light curve of the form
\bea
L_B(t) = f_B \, f_{\rm Edd} \, L_{\rm Edd}\exp{\Bigl(-\frac{t}{t_{\rm Q}}\Bigr)} \,,
\eea
where $f_{\rm Edd}$ is the ratio of the bolometric luminosity to the 
Eddington luminosity, and $f_B$ is the fraction of the bolometric 
luminosity radiated in the $B$-band.
 
%
%
\begin{figure}[tbh]
\centerline{\includegraphics[width=\textwidth]{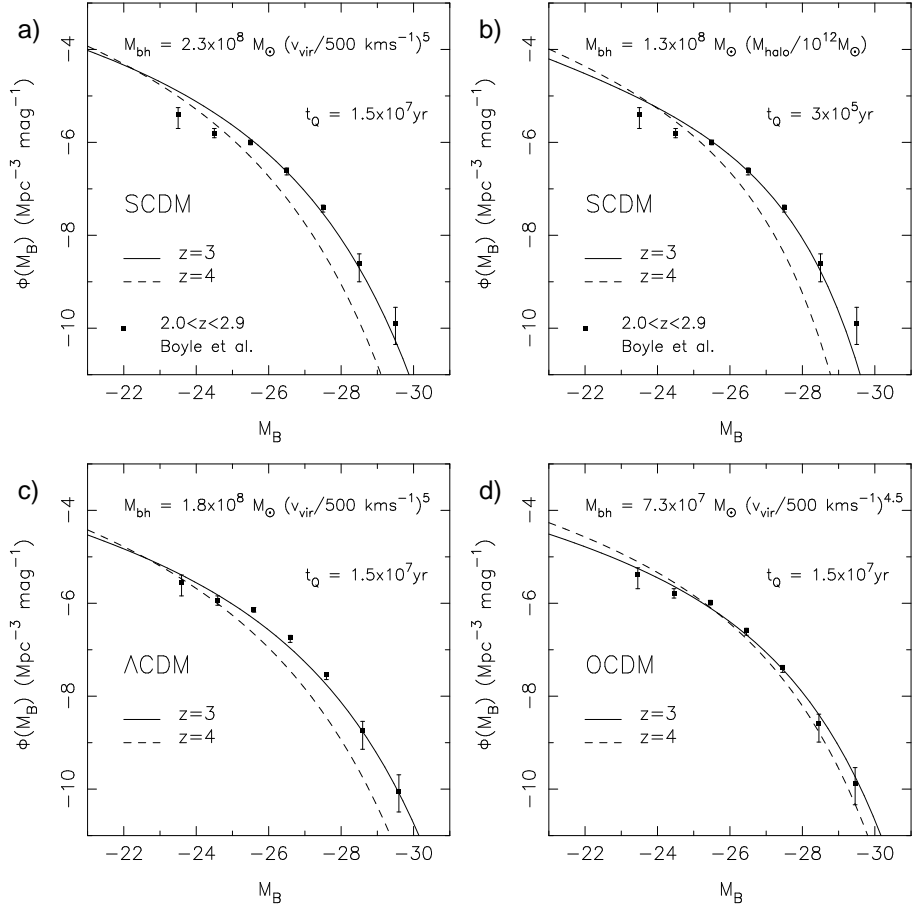}}
\caption{$B$-band quasar LF at $z = 3$ for three
cosmological models, computed using the time derivative of the space
density of DM halos. A factor of 6.0 bolometric correction factor has 
been applied. Panels (a), (c), and (d) assume a non-linear relation 
between the black hole mass and the halo mass with a quasar lifetime 
of $1\times 10^7$~yr, and panel (b) assumes a linear 
relation between the accreting black hole mass and the DM halo mass 
with a shorter lifetime of $10^6$~yr for the quasar
[the same parameters as the best fit model explored in Haiman 
\& Loeb 1998]. Panels (a) and (b) are for the SCDM model, 
and (c) and (d) are for the $\Lambda$CDM model and the 
OCDM model, respectively. For details of the parameters 
characterizing the variants of CDM plotted here, see 
Haehnelt et al.\ (1998).
Over-plotted data points are from Boyle, Shanks, \& Peterson (1988).}
\label{priyafig2}
\end{figure}
 
Given the number of free parameters, this is only a consistency
check, but we find that with a simple set of assumptions, we are 
able to connect the CDM mass functions to a model of optically 
bright quasars that the DM halos can host. Figure~\ref{priyafig2} 
shows how---if we allow ourselves some freedom in the relation 
between halo mass and black hole mass---reasonable fits are obtained 
for a range of lifetimes and for all of the CDM variants 
(the variants considered are the Standard CDM [SCDM]
model with $\Omega_{\rm tot} = 1$ and $\Omega_{\Lambda} = 0$, the Open CDM
[OCDM] model with $\Omega_{\rm tot} = 0.3$ and $\Omega_{\Lambda} = 0$, 
and the Lambda CDM [$\Lambda$CDM] model with $\Omega_{\rm tot} = 1$ 
and $\Omega_{\Lambda} = 0.7$). There are, however,
systematic trends: with increasing lifetime, (1) the black hole mass 
has to become a progressively more non-linear function of the halo mass, 
and (2) the black hole formation efficiency has to decrease in order to 
match the LF of quasars. This is due to the fact that quasars are
identified with rarer and more massive halos with increasing lifetime,
and these fall on successively steeper portions of the halo mass
function. 

In three of the panels in Figure~\ref{priyafig2}, we assume a quasar 
lifetime close to the Salpeter timescale and the following scaling of 
the black hole mass with the halo virial velocity:
\bea 
\mh\propto\, v_{\rm halo}^{5}\propto M_{\rm halo}^{5/3} (1+z)^{5/2} \,.
\eea 
This particular choice of dependence is in very good agreement with 
the tight empirical correlation found between the black hole mass and 
the velocity dispersion of the host bulge for local galaxies
(Ferrarese \& Merritt 2000; Gebhardt et al.\ 2000; Tremaine et al.\ 2002). 
The remaining panel (b) shows a linear relation between the halo 
mass and the black hole mass, as advocated by Haiman \& Loeb (1998); this 
requires a quasar lifetime of less than $10^{6}$~yr---much shorter 
than the Salpeter time for the usually assumed values of $\epsilon$. 

In principle, $t_{Q}$ could also depend on mass or other
parameters. The main uncertainty is the duty cycle or lifetime of
the optically bright phase of the quasar. The lifetime is degenerate with
the fraction of galaxy halos that host a black hole. Menou, Haiman, \&
Narayanan (2001) showed that, despite the ubiquity of SMBHs
at the centers of luminous nearby galaxies, only a small
fraction need to harbor SMBHs at high redshifts,
due to the assembly of structure at late times via multiple
mergers. This fact needs to be taken into account while studying the
clustering properties of quasars. It is precisely those 
observations---clustering as a function of redshift---that will help 
constrain the lifetime and hence break the degeneracy. 

We should also note that the
lifetime of the optically bright quasar phase probably reflects only
a small fraction of the time that the black hole is accreting
gas. Direct evidence for this comes from X-ray surveys, which show
that about 4\% of the $>L_*$ galaxy population are X-ray luminous
(Barger et al.\ 2001a, 2001b). This implies that accretion at a level
sufficient to produce detectable X-ray emission lasts for at least
half a gigayear in these galaxies.
 
\section{The Semi-Analytic Approach}

Semi-analytic models of galaxy formation track the formation and 
evolution of galaxies within a merging hierarchy of dark matter 
halos in a CDM universe (e.g., Kauffmann \& White 1993). Simple 
recipes are adopted to describe gas cooling within these halos, 
star formation, supernova feedback, and merging rates of galaxies. 
Stellar population synthesis models are used to generate galaxy
LFs, counts, and redshift distributions for direct comparison
with observations. These models have been studied extensively and 
can reproduce many properties of observed galaxies, including
variations in galaxy clustering with morphology, luminosity, and 
redshift (Baugh et al.\ 1998; Kauffmann et al.\ 1999),
the evolution of cluster galaxies (Kauffmann \& Charlot 1998), 
and the properties of the LBG population 
(Baugh et al.\ 1998; Somerville, Primack, \& Faber 2001). 

Into this merging scheme, Kauffmann \& Haehnelt (2000) incorporated
the growth of SMBHs. It is assumed that SMBHs are fueled with gas 
and hence grow via 
accretion during major mergers. In addition, it is assumed that when 
two galaxies of comparable mass merge, their central black holes
also merge, accompanied by gas accretion over the Salpeter time. 
With these simple assumptions, the model proposed by 
Kauffmann \& Haehnelt (2000, 2002) fits many aspects of the 
evolution of galaxies. It successfully reproduces (1) the observed 
correlation between bulge luminosity and black hole mass 
(Magorrian et al.\ 1998; Gebhardt et al.\ 2000), and the tighter
version of the same---the velocity dispersion of the bulge versus 
black hole mass in nearby galaxies (Ferrarese \& Merritt 2000; 
Gebhardt et al.\ 2000; Tremaine et al.\ 2002), (2) what is
currently known of the evolution of the abundance of quasars with 
redshift, and (3) the relation between quasars and their host galaxy 
luminosities. Kauffmann \& Haehnelt (2000) demonstrate that the 
sharp decline observed in the number density of quasars from 
$z \sim 2$ to $z = 0$ reflects a combination of effects: a decrease 
in the merging rate of DM halos in this redshift range, depletion 
in the availability of cold gas to fuel these black holes, and an 
increased timescale for gas accretion.   
 
\section{Constraints on the Accretion History
of Supermassive Black Holes}
 
The question of when SMBHs gained most of their
mass is closely related to $t_{\rm Q}$ and $f_{\rm Edd}$.
For bright quasars, $f_{\rm Edd}$ must be $> 0.1$; otherwise,
excessively massive individual black holes would be required to
explain the most luminous quasars detected at high redshifts. If
$\epsilon = 0.1$, then significant growth could not have occurred in a
low radiative efficiency phase, either after the optically bright
phase (in which case it would be in an advection dominated accretion
flow or ADAF), or as a result of over-fed accretion during the
optically bright phase. Furthermore, $f_{\rm Edd}$ will always be
smaller than unity, even if the ratio of the accretion rate to that
necessary to sustain the Eddington luminosity, $\dot m$, greatly
exceeds unity (over-fed accretion case). This is because a ``trapping
surface'' develops at a radius proportional to $\dot m$ in the
accretion disk, within which the radiation advects inwards rather than
escapes.  In consequence, the emission efficiency declines inversely
with $\dot m$ for $\dot m >1$ (Begelman 1978).

If $\epsilon$ exceeds 10\%, then the inferred mass density gained
during the optically bright phase is proportionally reduced, allowing
room for a significant fraction of the mass to be accreted in an
obscured phase. This ties in very nicely with the large number of
obscured sources detected in deep X-ray (e.g., Barger et al.\ 2003)
and submillimeter (see Blain et al.\ 2002 for a review) surveys
and is a very plausible scenario, at least if the typical 
black hole has a non-zero spin. A significant spin is more likely 
to arise as a result of rapid accretion rather than growth via black 
hole mergers (Hughes \& Blandford 2003; see also 
Gammie, Shapiro, \& McKinney 2003).

%
%
\begin{figure}[tbh]
\centerline{\includegraphics[width=\textwidth]{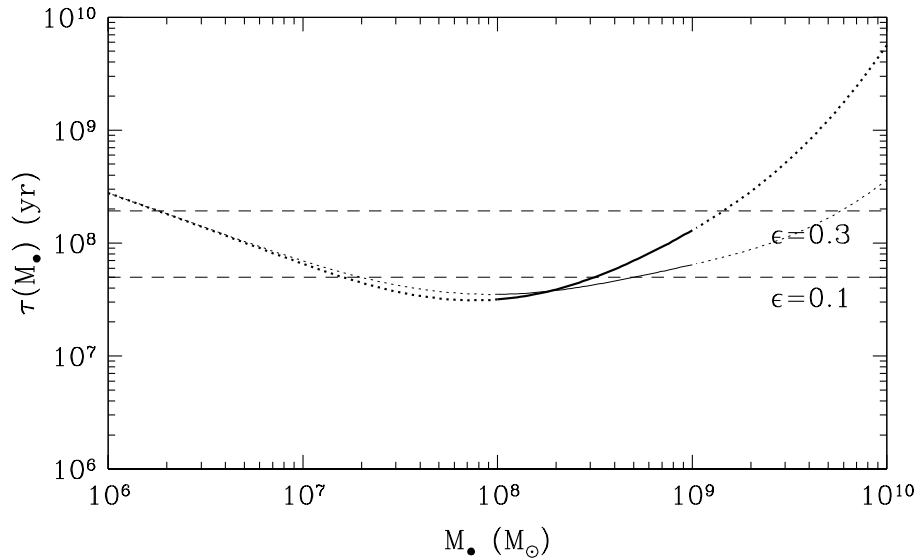}}
\caption{Estimate of the mean lifetime of quasars.
Reliable constraints are obtained for the mass range
$\mh=10^{8-9}~\msun$. Figure from Yu \& Tremaine (2002; their Fig.~5).}
\label{priyafig3}
\end{figure}

Using the observed quasar LF, the observed black hole
mass function in local early-type galaxies at $z = 0$, and assuming
that quasars are accreting at the Eddington limit, Yu \& Tremaine (2002)
derive constraints on the mean lifetime of quasars (i.e., the duration
of their optically bright phase; Fig.~\ref{priyafig3}). 
The lifetime obtained as a function of black hole mass is $\sim\,3-13\,
\times\,10^7$ yr. Thus, current observations are consistent with
the fact that the primary growth phase for black holes has a duration
comparable to the Salpeter time, implying
that accretion during the optically bright quasar phase can
add substantial mass to the black holes (as opposed to mass accretion
via mergers alone). To avoid over-producing the mass in local
black holes, the mean accretion rate subsequent to the optically
bright phase must be more than two orders of magnitude lower than the
Eddington rate. Theoretically, this requires a relatively low feeding
rate within ADAF models, in which the mass supply at large radius is
equal to the accretion rate at the event horizon (Narayan \& Yi 1995).
No such constraint is obtained if the low radiative efficiency is
primarily due to mass loss from accretion flows (Blandford \& Begelman
1999).

\section{Possible Accretion Histories}

%
%
\begin{figure}[tbh]
\centerline{\includegraphics[width=0.4\textwidth]{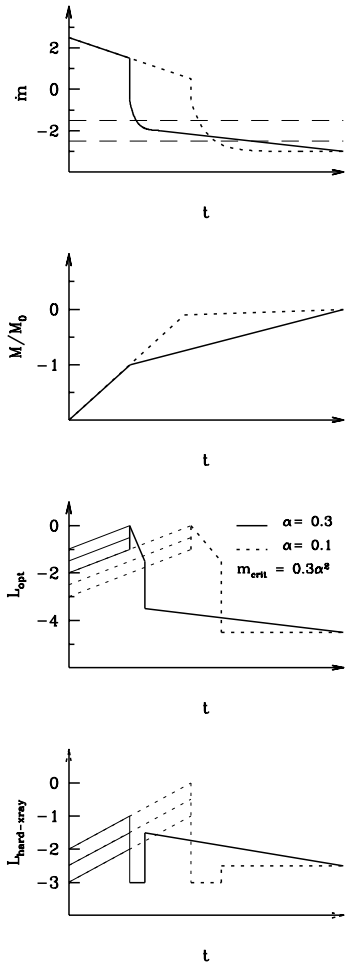}}
\caption{Two accretion histories with low overall optical
emission efficiencies for producing blue light
({\em solid curves\/}---most of the mass is accreted during a late
and prolonged ADAF phase; {\em dashed curves\/}---black hole gains 
most of its mass during a short lived early phase with $\dot m>1$).
Panels show {\em (from top to bottom)\/} mass accretion rate in terms
of Eddington accretion rate, mass relative to
final mass, and optical and hard X-ray luminosities.
Dashed lines in top panel indicate the critical
accretion rate for $\alpha=0.1$ and $\alpha=0.3$, where $\alpha$ is
the disk viscosity.
SED for accretion with
$\dot m>1$ is rather uncertain (as indicated by the three
parallel lines for $\dot m>1$ in the two bottom panels)
and should depend on the absorbing column and the dust content
of the outer parts of the self-gravitating disk and the host galaxy.
Figure from Haenelt et al.\ (1998; their Fig.~4).}
\label{priyafig4}
\end{figure}

In Figure~\ref{priyafig4}, we illustrate the observable effects of 
two possible accretion histories with low overall efficiencies for 
producing blue light---the solid curves describe an accretion history 
where most of the mass is accreted during the ADAF phase, while the 
dashed curves describe an accretion history where the black hole 
gains most of its mass during a
short lived early phase with $\dot m>1$. We show versus time 
{\em (top panel)\/} the mass accretion rate in units of the 
Eddington accretion rate, {\em (second panel)\/} the mass relative 
to the final mass, and {\em (bottom two panels)\/} the optical and 
hard X-ray luminosities.
The accretion rate is constant at the beginning with $\dot
m >1$. The mass therefore linearly rises and $\dot m$
decreases. The spectral energy distribution (SED) for accretion with 
$\dot m>1$ is rather uncertain (as indicated by the three parallel 
lines for $\dot m > 1$ in the two bottom panels) and should depend on 
the absorbing column and the dust content of the outer parts of the
self-gravitating disk and/or the host galaxy. The sharp drop of 
$\dot m$ marks the onset of the back-reaction on the accretion flow, 
and either the start or the peak of the optically bright phase (with a
rather inefficient production of hard X-rays). Once the accretion
rate has fallen below the critical rate for an ADAF (indicated by the
dashed lines in the top panel), the SED will
change to one peaked in the hard X-ray waveband.

\section{Faint X-ray Sources and the Hard X-ray Background}
 
The X-ray emission from quasars provides direct information about
their black hole regions and where accretion occurs, as the X-rays
are produced by the inner accretion disk and its corona. X-ray
observations can address the fundamental question of whether
early black holes at high redshifts grow and feed in a similar
fashion to their low redshift counterparts. The comoving
number density of quasars declines dramatically over cosmic time
(Fig.~\ref{priyafig1}), and part of this
strong evolution is believed to be due to changes in the environment
that are likely to impact the X-ray emission region 
(Vignali et al.\ 2003).

As pointed out by many authors, the X-ray emission of optically
selected quasars is too soft to explain the origin of the hard X-ray
background. Di Matteo \& Fabian (1997) and Yi \& Boughn (1998) argued
that the emission from ADAFs has a spectral shape similar to the hard
X-ray background. Fabian et al.\ (1998) suggested that this might 
also be true for dust obscured accretion. The {\it ROSAT}, 
{\it Chandra}, and {\it XMM-Newton} X-ray observations and 
ground-based follow-ups (e.g., Almaini et al.\ 1996; 
Hasinger et al.\ 1998, 2001; Schmidt et al.\ 1998; McHardy et al.\ 1998; 
Mushotzky et al.\ 2000; Brandt et al.\ 2001; 
Barger et al.\ 2001b, 2002; Giacconi et al.\ 2002; 
Szokoly et al.\ 2004) that resolve
the X-ray background into discrete sources show that obscured AGN are
the primary contributors to the X-ray background. The majority of
these sources are at low redshifts; thus, while they contribute
to the integrated black hole mass density, their significance compared
to the optically bright quasars is uncertain (see Cowie \& Barger,
this volume, for a discussion of this issue). That accretion occurs in
galaxies other than luminous AGN also fits in well with the detection
of extremely low level optical AGN activity in a large fraction of
galaxies reported by Ho, Filippenko, \& Sargent (1997).

\section{Conclusions}
 
The optical quasar LF at $z\sim 3$ can be plausibly matched with 
the LF of star-forming galaxies at the same redshift, and with 
the mass function of DM halos predicted by a range of variants of 
CDM cosmogonies (and believed to comply with observational
constraints in the low redshift universe). This is possible for
lifetimes of optically bright quasars in the range $10^{7}$ to
$10^{8}$~yr. There is a correlation between the lifetime and the
required degree of non-linearity in the relation between black hole
and halo mass. The non-linearity has to increase for increasing
lifetime. Predicted host halo masses, host galaxy luminosities, and
the clustering strength all increase with increasing lifetime, and
further observations of these offer our best hope of constraining 
the duration of the optically bright phase of quasars.
 
The present-day black hole mass density implied by the integrated
luminosities of optically bright quasars is comparable to that inferred
from recent black hole estimates in nearby galaxies for generally
assumed efficiencies for producing blue light. This limits the 
contribution of accretion in a low radiative efficiency mode to
either (1) modest rates at low redshifts or (2) highly 
super-Eddington rates during the final assembly of black holes.

While the conventional picture of quasars as a population of
supermassive black holes growing by accretion and mergers seems more
secure than ever, there remain many open key questions. The importance
of several parameters is as yet poorly understood: the role of the
black hole mass, the accretion rate, the radiative efficiency, the 
properties of accretion flows at low and high accretion rates, 
the relative importance of mergers versus accretion as a function 
of cosmic epoch, and the relation between quasar populations observed 
at different wavelengths. Bearing in mind that fundamental quantities
like the radiative efficiency might well be functions of the black
hole mass, redshift, and environment---rather than constants, as assumed
hitherto for simplicity---it is clear that the comparison of integrated
quantities alone is insufficient to uniquely specify the model.

Recent modeling attempts have, however, made considerable progress in
integrating multiwavelength data into the schemes outlined here
(Steed \& Weinberg 2004; Yu \& Tremaine 2002). The key observations
in the future that are likely to constrain theoretical models are
probing the faint end of the quasar LF, determining the
clustering properties of quasars with redshift, and obtaining better 
measurements of the joint X-ray, optical, infrared, and submillimeter 
luminosity functions.

{}

\end{document}